 \documentclass[iop]{emulateapj}   \usepackage{apjfonts}

\usepackage{natbib}
\usepackage{epsf}
\usepackage{color}
\usepackage{amsmath}

\newcommand{\spitzer}{\textit{Spitzer}}

\newcommand{\lsim}{\lesssim}
\newcommand{\gsim}{\gtrsim}

\newcommand{\eg}{e.g.}

\newcommand{\msol}{\hbox{$M_\odot$}}

\newcommand{\mone}{\hbox{$[3.6]$}}
\newcommand{\mtwo}{\hbox{$[4.5]$}}
\newcommand{\mthree}{\hbox{$[5.8]$}}
\newcommand{\mfour}{\hbox{$[8.0]$}}

\newcommand{\fulltarget}{ClG~J0218.3-0510}
\newcommand{\target}{ClG~J0218.3-0510}

\shorttitle{SPITZER--SELECTED CLUSTER AT $z=1.62$}
\shortauthors{PAPOVICH ET AL.}

\begin{document}

\slugcomment{Accepted for Publication in the Astrophysical Journal}
\title{A Spitzer--Selected Galaxy Cluster at $z$=1.62\altaffilmark{1}}

\author{\sc C.~Papovich\altaffilmark{2}, 
  I.~Momcheva\altaffilmark{3}, 
  C.~N.~A.~Willmer\altaffilmark{4}, 
  K.~D.~Finkelstein\altaffilmark{2}, 
  S.~L.~Finkelstein\altaffilmark{2}, 
 K.--V.~Tran\altaffilmark{2},
  M.~Brodwin\altaffilmark{5}, 
  J.~S.~Dunlop\altaffilmark{6}, 
  D.~Farrah\altaffilmark{7}, 
  S.~A.~Khan\altaffilmark{8,9}, 
  J.~Lotz\altaffilmark{10},
  P.~McCarthy\altaffilmark{3}, 
  R.~J.~McLure\altaffilmark{6},
  M.~Rieke\altaffilmark{4},
  G.~Rudnick\altaffilmark{11}
  S.~Sivanandam\altaffilmark{4},
 F.~Pacaud\altaffilmark{12}, and
  M.~Pierre\altaffilmark{13}
}
\altaffiltext{1}{ This work is based in part on observations made with the Spitzer Space Telescope, which is operated by the Jet Propulsion laboratory, California Institute of Technology, under NASA contract 1407.  This paper also includes data
gathered with the 6.5 meter Magellan Telescopes located at Las
Campanas Observatory, Chile.  This work is based in part on data collected at Subaru Telescope, which is operated by the National Astronomical Observatory of Japan.}
\altaffiltext{2}{George P.\ and Cynthia Woods Mitchell Institute for Fundamental Physics and Astronomy, and 
Department of Physics and Astronomy, Texas A\&M University, College Station, TX, 77843-4242; papovich@physics.tamu.edu}
\altaffiltext{3}{Observatories, Carnegie Institution of Washington, 813 Santa Barbara St., Pasadena, CA, 91101}
\altaffiltext{4}{Steward Observatory, University of Arizona, 933 N.~Cherry Ave., Tucson, AZ 85721}
\altaffiltext{5}{W. M. Keck Postdoctoral Fellow, Harvard--Smithsonian Center for Astrophysics, 60 Garden St., Cambridge, MA 02138}
\altaffiltext{6}{Institute for Astronomy, Royal Observatory, University of Edinburgh, UK}
\altaffiltext{7}{Astronomy Centre, University of Sussex, Falmer, Brighton, UK}
\altaffiltext{8}{
Pontificia Universidad Cat\'olica, Departamento de Astronom\'ia y
Astrof\'isica, 4860 Av. Vicu\~na Mackenna, Casilla 306, Santiago 22 Chile}
\altaffiltext{9}{
Shanghai Key Lab for Astrophysics, Shanghai Normal University, Shanghai
200234, China}
\altaffiltext{10}{Leo Goldberg Fellow, National Optical Astronomy Observatories, 950 N.~Cherry Ave., Tucson, AZ 85719}
\altaffiltext{11}{Department of Physics and Astronomy, University of Kansas, 1251 Wescoe Hall Dr., Lawrence, KS, 66045-7582}
\altaffiltext{12}{Argelander Institute for Astronomy, Bonn University,
Auf dem H\"ugel 71, 53121 Bonn, Germany}
\altaffiltext{13}{Service d'Astrophysique, CEA Saclay, 91191 Gif sur
  Yvette, France}



\begin{abstract}  

\noindent We report the discovery of a galaxy cluster at $z$=1.62
located in the \spitzer\ Wide-Area Infrared Extragalactic survey
XMM-LSS field.  This structure was selected solely as an overdensity
of galaxies with red \spitzer/IRAC colors, satisfying
$(\mone-\mtwo)_\mathrm{AB}>-0.1$ mag.  Photometric redshifts derived
from Subaru XMM Deep Survey ($BViz$-bands), UKIRT Infrared Deep
Survey--Ultra-Deep Survey (UKIDSS-UDS, $JK$-bands), and from the
\spitzer\ Public UDS survey (3.6-8.0~\micron) show that this cluster
corresponds to a surface density of galaxies at $z\approx 1.6$ that is
$>20\sigma$ above the mean at this redshift.      We obtained optical
spectroscopic observations of galaxies in the cluster region using
IMACS on the Magellan telescope.  We measured redshifts for seven
galaxies in the range $z$=1.62--1.63  within 2.8 arcmin ($<1.4$~Mpc)
of the astrometric center of the cluster.   A posteriori analysis of the
\textit{XMM} data in this field reveal a  weak ($4\sigma$) detection
in the [0.5--2~keV] band compatible with
the expected thermal emission from such a cluster.  The
color--magnitude diagram of the galaxies in this cluster shows a
prominent red-sequence, dominated by a population of red galaxies with
$(z-J)>1.7$ mag.  The photometric redshift probability distributions
for the red galaxies are strongly peaked at $z=1.62$, coincident with
the spectroscopically confirmed galaxies. The rest--frame $(U-B)$
color and scatter of galaxies on the red-sequence are consistent with
a mean luminosity--weighted age of $1.2\pm0.1$ Gyr, yielding a
formation redshift $\overline{z_f}=2.35\pm0.10$,  and corresponding to
the last significant star-formation period in these galaxies.   
%
%
\end{abstract}
 


\section{INTRODUCTION}

Galaxy clusters provide important samples to study both the evolution
of large--scale structure and the formation of galaxies.   The
evolution of the number density of massive galaxy clusters involves
primarily gravitational physics, which depends strongly on the  cosmic
mass density, the normalization and shape of  the initial power
spectrum, as well as on the dark energy equation of state
\citep[\eg,][]{eke98,bahcall99,borg01,haim01,spri05a,pacaud07,jee09}. 
%
%
Because the massive galaxies in clusters formed nearly
contemporaneously at $z \gg 1$ with similar star--formation and
assembly histories \citep[\eg,][]{stan98,tran07,eise08}, observations
of the evolution of distant cluster galaxies provide strong constraints on
hierarchical galaxy evolution models, which make detailed predictions
for the formation of these objects \citep[\eg,][]{delucia07a}.
%
%


Studies of high--redshift clusters have been frustrated by
small sample sizes.   Few clusters have confirmed redshifts beyond $z
\sim 1.3$ \citep[\eg,][]{mullis05,stan05,brod06,stan06,eise08,
kurk09,wilson09}.  This dearth of detected galaxy clusters at
$z > 1.3$ stems from the significant challenges and potential biases
in identifying these structures. Deep X--ray surveys have identified
spectroscopically confirmed clusters to $z \lsim 1.5$
\citep[\eg,][]{rosati04,stan06}, but X--ray selection typically require relaxed
systems, which is unlikely to be the case at high redshifts where
cluster progenitors will be less massive and more disordered, with
less time for the intracluster medium (ICM) to thermalize
\citep[\eg,][]{rosati02}.   Searches for galaxy overdensities
around distant radio galaxies require the presence of a massive
central galaxy
\citep[\eg,][]{kurk00,miley04,stern03,koda07,vene07,zirm08,chia10},
which may not be an intrinsic property of cluster progenitors.  

Other searches utilize the empirically observed, tight
color--magnitude relation in central cluster galaxies \citep[the ``red
sequence'', e.g.,][]{visv77,glad05,glad07,kaji06,muzzin09a}.   However, this
selection is biased  potentially  against clusters whose galaxies have
had recent star formation \citep[and overdensities of blue,
star--forming galaxies at high redshift have been identified,
e.g.,][]{stei05}.    Most studies of cluster galaxies imply their
stellar populations formed at $z_f \gsim 1.5$
\citep[\eg,][]{vandokkum07a}, concurrent with the peak epoch in the
star--formation rate density from UV and IR measurements
\citep[see \eg,][]{hopk06}.  As searches for clusters approach
the redshift of their formation, cluster galaxies should show
increasing indications for star formation with less time available to
build up a substantial red--sequence population. 

We have initiated a search for galaxy cluster candidates at $z > 1.3$
selected solely as  overdensities of galaxies with red $\mone - \mtwo$
colors using data from the Infrared Array Camera
\citep[IRAC,][]{fazio04}  on board \spitzer\ \citep{werner04}
following the method of \citet{papo08}.\footnote{Throughout we denote magnitudes measured in
the 3.6, 4.5, 5.8, and 7.9~\micron\ IRAC channels as \mone, \mtwo,
\mthree, and \mfour, respectively.}   At $z$$<$1 model stellar
populations have blue $\mone - \mtwo$ colors because these bands probe
the stellar Rayleigh--Jeans tail  \citep[with the expection of some
IR--luminous, star--forming galaxies at $z\sim 0.3$ which have a
contribution of warm dust to their near--IR colors, see][]{papo08}.  At
$z$$\gsim$1 both star--forming and passively  evolving stellar
populations appear red in $\mone - \mtwo$ as these bands probe the
peak of the stellar emission at 1.6~\micron\
\citep[see][]{simp99,sawi02,papo08}.  Therefore, selecting
overdensities of red $\mone - \mtwo$ sources potentially identifies
high--redshift cluster candidates with little bias from galaxy stellar
populations.    \citet{papo08} showed that IRAC--selected $z > 1.3$
cluster candidates from the \spitzer\ Wide--Infrared Extragalactic
(SWIRE) survey have   clustering scale lengths of $r_0 \approx
20$~$h^{-1}$ Mpc, consistent with other high--redshift galaxy clusters
\citep[see][]{brod07}. 

Here we report the discovery and spectroscopic confirmation of a
galaxy cluster, \fulltarget,  \citep[corresponding to \textit{Infrared
  Cluster} ``A'', IRC-0218A, 
in the 0218-051 field, selected in][]{papo08},
and we discuss its photometric and spectroscopic properties.  This
galaxy cluster was identified using our IRAC color selection with no
other additional criteria imposed.    
%
%
In \S~2 we discuss the target selection, and photometric and
spectroscopic observations.  In \S~3 we present the evidence
supporting the assertion that this structure is a galaxy
cluster, and in \S~4 we discuss its properties.  In \S~5 we summarize
our results.    Unless otherwise noted we report all magnitudes in
reference to the \citet{john66} magnitude system relative to Vega.  We
explicitly denote magnitudes relative to the  absolute bolometric
system \citep{oke83} with an AB subscript, $m_\mathrm{AB} = 23.9 - 2.5
\log (f_\nu / \mathrm{1\,\mu Jy})$.  We use cosmological parameters
$\Omega_m=0.3$, $\Lambda=0.7$, and $H = 70$ km s$^{-1}$ Mpc$^{-1}$
throughout.  For this cosmology, the angular diameter distance is
$0.5$~Mpc arcmin$^{-1}$ at $z=1.62$.

\section{SAMPLE SELECTION AND SPECTROSCOPIC OBSERVATIONS}

\subsection{Identification of High--redshift Galaxy Cluster Candidates}

\begin{figure*}
\epsscale{1.17}  
\plottwo{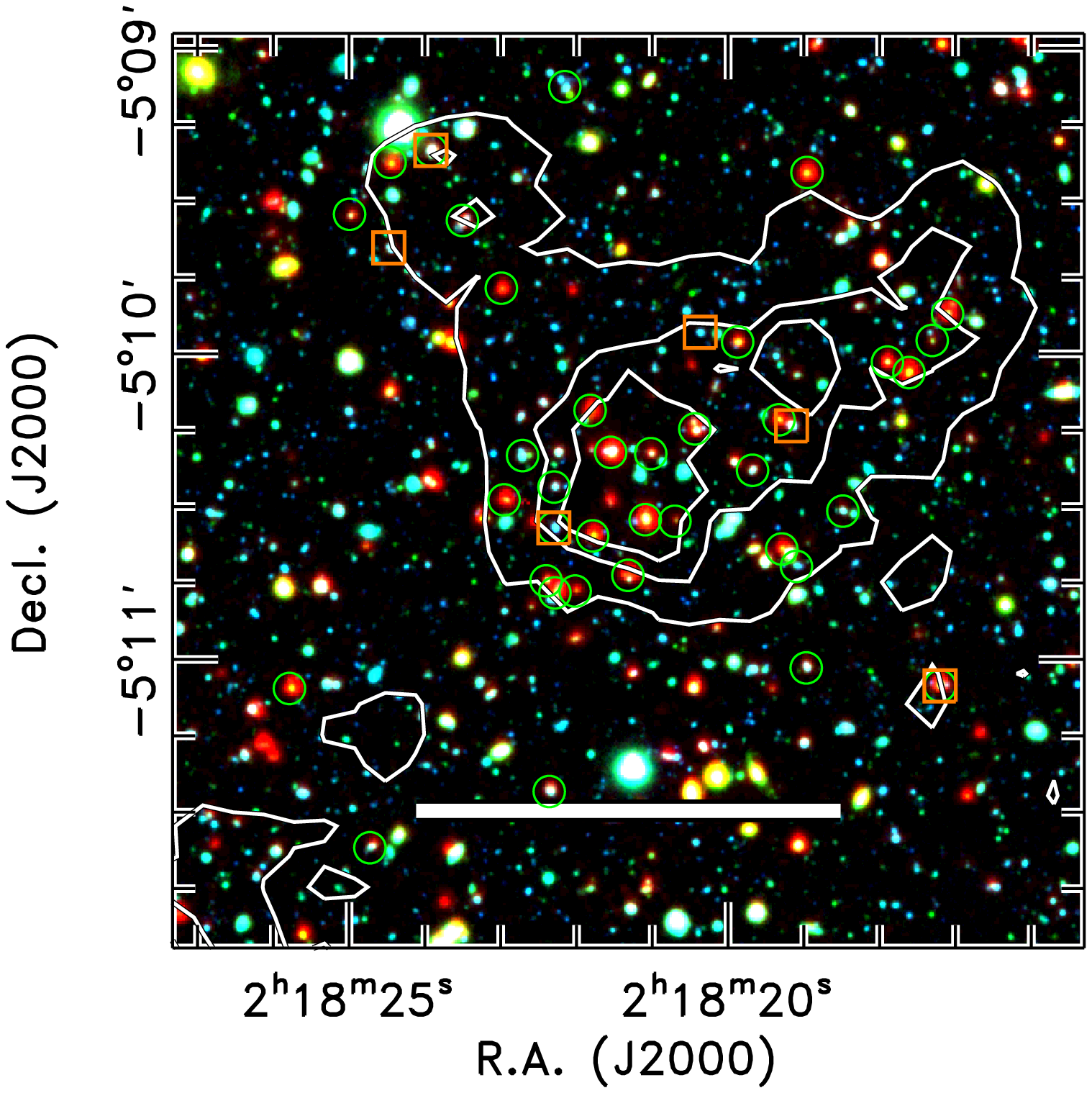}{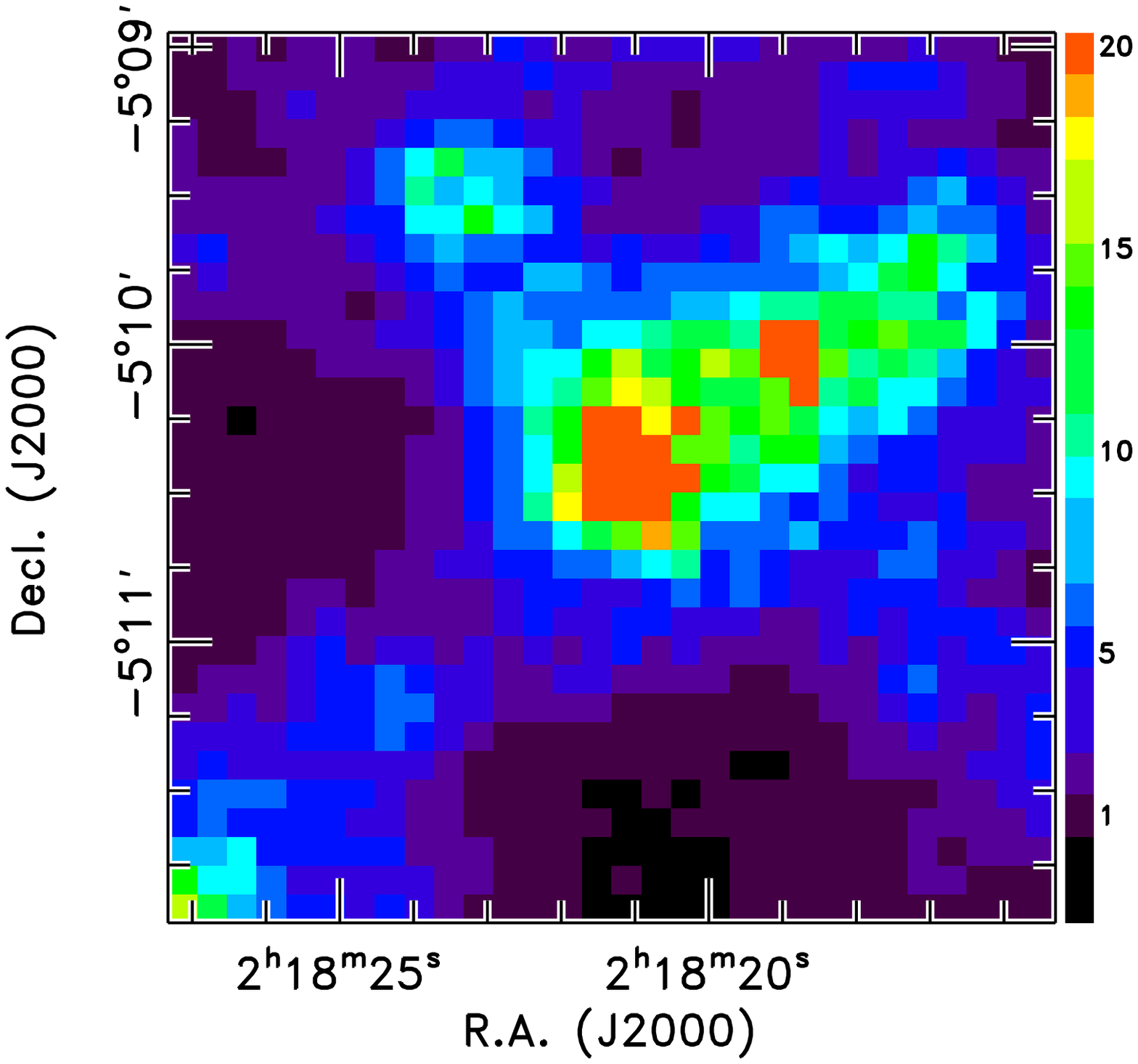}
\epsscale{1.0}  
\caption{ The left panel shows a false--color image of the target
field.   Blue corresponds to the Suprime--Cam $B$--band, Green to the
Suprime--Cam $i$--band, and Red to the \spitzer\ 4.5~\micron\ band.   The
images have not been corrected for variations in the data image
quality.   The image spans $3\arcmin \times 3\arcmin$, corresponding
to $1.5\, \mathrm{Mpc} \times 1.5\, \mathrm{Mpc}$ at
$z=1.62$.   The heavy white bar shows a distance of $0.7$~Mpc
at the redshift of the cluster.  Small green circles denote candidate
cluster members with $\mathcal{P}_z > 0.5$ as defined in
\S~\ref{section:photoz}.   Orange squares denote those objects with
spectroscopically confirmed redshifts $1.62 < z < 1.65$.   The
contours denote regions with 5, 10, and 15$\sigma$ above the mean
density of galaxies with $1.5 < z_\mathrm{phot} < 1.7$.  The right
panel shows the surface density of galaxies with $1.5 <
z_\mathrm{phot} < 1.7$ in units of the number of standard deviations
($\sigma$) above the mean density, ranging from -1 to 20 (as indicated
in the plot legend).   \label{fig:rgb}}
\end{figure*}

We identified galaxy cluster candidates at $z\gsim 1.3$ using the IRAC
data from the SWIRE survey.  These data cover roughly 50 deg$^2$
divided over six fields separated on the sky \citep{lons03}.  We used a
simple color selection to identify high--redshift galaxies from
\spitzer/IRAC 3.6 and 4.5~\micron\ photometry.   As discussed in
\citet{papo08} $>$90\% of galaxies with $z > 1.3$ have
$(\mone - \mtwo)_\mathrm{AB} > -0.1$ mag.  \citet{papo08}
identified candidate galaxy clusters at $z > 1.3$ by selecting
overdensities of $\gsim 30$ objects satisfying this color criterion
within radii of $1\farcm4$ (corresponding to $r < 0.7$ Mpc at
$z=1.5$) from the SWIRE survey data.   
%
%
\target\ was identified in \citet{papo08} in the SWIRE XMM-LSS field,
and has astrometric coordinates, $\alpha = 2^h18^m21.3^s$, $\delta =
-05^\circ10^\prime27^{\prime\prime}$ (J2000), derived from the
centroid of the SWIRE IRAC sources in this overdensity.
%
%
The left panel of figure~\ref{fig:rgb} shows a false--color image
using the $B$--band (blue), $i^\prime$--band (green) and 4.5~\micron\
image (red) of the field centered on the coordinates of the
IRAC--selected overdensity.

\subsection{Optical and Near--IR Imaging}

The XMM-LSS includes deep optical (0.4--1~\micron) and near--IR
imaging (1--2~\micron) in a portion of the field covering roughly 0.70
deg$^2$.   Optical imaging is available from the Subaru--XMM Deep
Survey \citep[SXDS,][]{furu08}, obtained with the Subaru
Prime Focus Camera (Suprime--Cam) in the broad bandpasses $B$, $R$,
$i^\prime$, and $z^\prime$.   Near--IR imaging is available from the
UKIRT Infrared Deep Sky Survey \citep[UKIDSS, data release
1,][]{lawr07} ultra deep survey (UDS) in the broad bandpasses $J$ and
$K$.   For the work here, we utilized the $K$--selected catalog of the
SXDS and UDS data from \citet{will09}.  These catalogs reach
$5\sigma$--limiting magnitudes in $1\farcs75$--diameter apertures of
$B_\mathrm{AB} < 27.7$, $R_\mathrm{AB} < 27.1$, $i_\mathrm{AB} <
26.8$, $z_\mathrm{AB} < 25.5$, $J_\mathrm{AB} < 23.9$, and
$K_\mathrm{AB} < 23.6$~mag.  The $K$--selected catalogs also include
quasi--total $K$ magnitudes, measured in elliptical apertures based on
the light profile for each source.   For details, we refer the reader
to \citet{will09}.    We corrected the fixed--aperture magnitudes to
quasi--total magnitudes using a unique aperture correction for each
source, defined as the difference between the fixed aperture and total
magnitude derived from the $K$--band for that source, 
$K(\mathrm{ap}) - K(\mathrm{tot})$.    We then applied the aperture
correction for each source to its fixed aperture magnitudes measured
in the other SXDF and UKIDSS bands. 

\subsection{\spitzer\ IRAC Imaging}

While the SWIRE IRAC data are sufficient for the selection and study
of the galaxies in this cluster, we made use of deeper IRAC imaging in
this field from the \spitzer\ public legacy survey of the UKIDSS UDS
(SpUDS, PI: J.~Dunlop).    The SpUDS data cover a field of area 1.5
deg$^2$ to deeper limiting flux densities than those available with
the SWIRE survey data.  We measured photometry in 4\arcsec--diameter
apertures in each of the IRAC 3.6, 4.5, 5.8, and 8.0~\micron\ images.
We applied correction factors of 0.32, 0.36, 0.55, 0.68~mag,
respectively, which corrects these aperture magnitudes to total
magnitudes for point sources.  We matched sources in each of the
IRAC--selected catalogs to those in the UKIDSS UDS $K$--selected
catalog  within 1\arcsec\ radii.  We combined the ``total''
aperture--corrected magnitudes for the IRAC data with the aperture
corrected photometry for the SXDS and UDS data.  

\subsection{Photometric Redshifts and the Integrated Redshift Probability}\label{section:photoz}

The merged SXDS, UDS, and IRAC catalog covers a wavelength baseline of
0.4--8~\micron, and we used these data to study the
cluster--candidates selected from the SWIRE IRAC data.  We derived
photometric redshifts for each source in the $K$--selected catalog
using EAZY \citep{bram08}.  We used the
default galaxy spectral energy distribution templates with a $K$--band
prior based on the luminosity functions of galaxies in a
semi--analytic simulation.  We derived the most likely photometric
redshift as well as the full photometric--redshift probability
distribution function, $P(z)$, normalized such that $\int\, P(z) dz =
1$ when integrated over all redshifts.   Our comparisons against the
spectroscopic redshifts in the SpUDS field \citep[C. Simpson, 
in preparation]{yama05,simp06,vanbreu07,vanbreu09}  showed that 
the most likely photometric redshifts have uncertainties of $\Delta([z_\mathrm{sp} - z_\mathrm{ph}] / [1 + z_\mathrm{sp}] ) = 0.04$ derived from the normalized median absolute deviation \citep{beers90} for
the more than 200 galaxies (excluding broad--line AGN) with
spectroscopic redshifts in the range $1.0 \le z \le 2.0$. 

To increase the efficiency of our spectroscopic observations, we
computed the surface density of galaxies in the SpUDS field in coarse
redshift intervals, and we prioritized those IRAC--selected cluster
candidates (selected over the much larger SWIRE field) that
corresponded also to large overdensities in photometric redshift.   To
measure the surface density of galaxies, we divided galaxies into
redshift intervals, $\Delta z = 0.2$, and measured the angular
distance from each object to the seventh--nearest neighbor, $d_7$, and
then computed the corresponding surface density, $\Sigma_7 \propto
(d_7)^{-2}$.  We tested other definitions for the nearest neighbor,
which produced similar results (changing the definition of the
$N$th--nearest neighbor changed primarily the angular resolution of
the surface density map).  We then calculated the mean and standard
deviation of the surface density across the entire UDS field.  

\target\ appears as a strong overdensity of galaxies with $1.5 <
z_\mathrm{ph} < 1.7$.  Figure~\ref{fig:rgb} (right panel) shows
the surface density of galaxies in this photometric--redshift range
centered on the IRAC--selected overdensity \target.  The
color shading corresponds to the number of standard deviations above
the mean surface density of galaxies at this redshift over the UDS
field.  \target\ corresponds to a $>20\sigma$ surface density of
galaxies in this redshift interval.  

Many of the galaxies in the field of \target\ have photometric
redshifts centered tightly around $z_\mathrm{ph} \simeq 1.6$.   We
quantified the likelihood of galaxies being associated in redshift by
defining the \textit{integrated redshift probability}, 
\begin{equation}
\mathcal{P}_z \equiv \int\, P(z)\, dz, 
\end{equation} 
integrated over $z=z_\mathrm{cen}\pm \delta z$.  For \target, we used
$z_\mathrm{cen}=1.625$ and $\delta z = 0.05 \times (1 +
z_\mathrm{cen})$ (therefore integrating over $1.49 < z < 1.76$),
approximately the 68\% confidence range on the photometric redshifts.
This was motivated by other cluster--member selection methods using
photometric--redshift--selected samples
\citep{brun00,hall04,eise08,pello09}.  Simply defined, the integrated
redshift probability is the fraction of the photometric redshift
probability distribution function within the redshift intervals
\citep[see discussion in][]{fink10}.  An integrated probability of
$\mathcal{P}_z = 0.5$ means that 50\% of the integrated photometric
redshift distribution lies between $z_\mathrm{cen} - \delta z < z <
z_\mathrm{cen} + \delta z$.   The galaxies in the field of \target\
with $\mathcal{P}_z > 0.5$ are indicated in figure~\ref{fig:rgb}.    
%

\begin{deluxetable*}{lccccccc}
\tablecaption{Spectroscopic Redshifts in the \target\ Field\label{table:specz}}
\tablecolumns{7}
\tablewidth{0pc}
\tablehead{
\colhead{R.A.} & \colhead{Decl.} & \colhead{$\mathcal{P}_z$} & \colhead{$z$} & \colhead{$\sigma_z$}  & \colhead{$R$} & \colhead{$\Delta$} & \colhead{$r$} \\ 
\colhead{(J2000)} & \colhead{(J2000)} & \colhead{} & \colhead{} & \colhead{} & \colhead{(mag)} & \colhead{(\arcmin)} & \colhead{(Mpc)}  \\
\colhead{(1)} & \colhead{(2)} & \colhead{(3)} & \colhead{(4)} &
\colhead{(5)} & \colhead{(6)} & \colhead{(7)} & \colhead{(8)}
}
\startdata
2$^h$18$^m$22.30$^s$  &      $-$5$^\circ$10$^{\prime}$34.7$^{\prime\prime}$  & 0.50 & 1.6224 & 0.0005 & 22.9 &  0.28 &  0.14  \\
2$^h$18$^m$20.37$^s$  &        $-$5$^\circ$09$^\prime$56.3$^{\prime\prime}$  & 0.42 & 1.6303 & 0.0012 & 23.6 &  0.57 &  0.29  \\
2$^h$18$^m$19.16$^s$  &      $-$5$^\circ$10$^{\prime}$14.7$^{\prime\prime}$  & 0.47 & 1.6230 & 0.0005 & 23.0 &  0.57 &  0.29  \\
2$^h$18$^m$24.14$^s$  &        $-$5$^\circ$09$^\prime$45.2$^{\prime\prime}$  & 0.46 & 1.5356 & 0.0004 & 23.2 &  0.99 &  0.51  \\
2$^h$18$^m$24.48$^s$  &        $-$5$^\circ$09$^\prime$39.7$^{\prime\prime}$  & 0.40 & 1.6228 & 0.0005 & 23.4 &  1.12 &  0.57  \\
2$^h$18$^m$17.20$^s$  &      $-$5$^\circ$11$^{\prime}$05.7$^{\prime\prime}$  & 0.61 & 1.6487 & 0.0006 & 23.8 &  1.21 &  0.61  \\
2$^h$18$^m$23.92$^s$  &        $-$5$^\circ$09$^\prime$20.6$^{\prime\prime}$  & 0.54 & 1.6222 & 0.0005 & 22.8 &  1.29 &  0.65  \\
 2$^h$18$^m$18.92$^s$  &        $-$5$^\circ$08$^\prime$00.3$^{\prime\prime}$  & 0.32 & 1.6234 & 0.0013 & 23.7 &  2.52 &  1.28  \\
2$^h$18$^m$17.46$^s$  &        $-$5$^\circ$08$^\prime$02.2$^{\prime\prime}$  & 0.48 & 1.4962 & 0.0003 & 22.9 &  2.60 &  1.32  \\
2$^h$18$^m$15.18$^s$  &        $-$5$^\circ$08$^\prime$11.9$^{\prime\prime}$  & 0.64 & 1.6224 & 0.0011 & 23.8 &  2.72 &  1.38  \\
2$^h$18$^m$14.50$^s$  &        $-$5$^\circ$06$^\prime$59.1$^{\prime\prime}$  & 0.47 & 1.6094 & 0.0008 & 23.5 &  3.86 &  1.96 
\enddata \tablecomments{(1) Right ascension, (2) Declination, (3)
Integrated redshift probability, (4) spectroscopic redshift, (5)
redshift uncertainty, (6) Suprime--Cam $R$--band magnitude, (7)
angular separation between the galaxy and the cluster astrometric
center (see \S~2.1), (8) projected physical separation between galaxy
and the cluster center for $z=1.62$.}
\end{deluxetable*}

\subsection{Spectroscopic Observations}\label{section:specz}

We targeted galaxies with high integrated redshift probability in the
region of \target\ using the \textit{Inamori Magellan Areal Camera and
Spectrograph} (IMACS) on the Magellan/Baade 6.5~m telescope on 2008
Oct 30--31, 2008 Nov 18--19, and again on 2009 Sep 11--12.  At the f/2
focus IMACS provides multiobject spectroscopic observations over a
$27.2\arcmin$--diameter field of view (FOV), which allowed us to
target roughly 100 galaxies in as many as 10 cluster candidates using
a single slitmask.   We targeted two separate IMACS fields within the
UDS, covering 10 cluster candidates in field 1 and 7 in field 2, using
two slitmasks per field to alleviate slit collisions.  \target\ was
one of our highest priority targets given the high surface density of
objects with high $\mathcal{P}_z$.   The other cluster candidates will
be discussed in a forthcoming paper. 

We observed with IMACS using the 200 lines/mm grating with the OG570
blocking filter, which provided $\sim$7~\AA\ resolution covering
$0.6-1$~\micron.   The IMACS CCDs were upgraded in 2008 and have very
high red sensitivity, with about a factor of two improvement in
sensitivity beyond 8500~\AA\ (A.\ Dressler 2008, private
communication).  We used the ``nod--and--shuffle'' mode
\citep{glaz01}  with $1\arcsec \times 2.2\arcsec$ slitlets, which
greatly improves the background subtraction and facilitated our
spectroscopic redshift success, especially for galaxies at $z > 1.3$.
For practical purposes, we prioritized galaxies with $R < 23.3$~mag,
but included galaxies to $R < 23.8$~mag (and a few even fainter
galaxies).  Conditions each night were generally clear, although not
photometric, and some time was lost to poor photometric conditions
(including most of the time during the 2009 Sept run).
Typical image quality during good conditions ranged over
$0.6-1$\arcsec\ at full width at half maximum for point sources.
Individual exposure times were 1800~s.  The total co-added exposure
times of data taken under good photometric conditions varied for each
mask, and were 3-4 hours on source.

We reduced the IMACS spectroscopic data using the Carnegie
Observatories System for MultiObject Spectroscopy (COSMOS,
v2.13)\footnote{http://obs.carnegiescience.edu/Code/cosmos}.  The
reduction steps to produce 2D spectra include wavelength calibration,
bias subtraction, flat fielding, sky subtraction, co--adding the
separate frames, and cosmic--ray removal.   We extracted 1D spectra at
each nod--and--shuffle position separately, as well as the associated
uncertainty on the spectra propagated through the reduction pipeline.
We then coadded the individually extracted 1D spectra for each target.
We used spectrophotometric standards taken at the end of each night to
provide flux calibration.   Because photometric conditions varied over
the course of each night, our photometric calibration is not absolute
(although this does not affect our ability to measure redshifts).  

We took spectra of  24 sources with $R < 23.8$~mag, 
integrated redshift probability $\mathcal{P}_z > 0.3$, and with an angular
separation of $\Delta < 4\arcmin$ of the astrometric centroid of
\target\ (corresponding to a physical separation of $r < 2.0$~Mpc at
$z=1.62$).  We measured spectroscopic redshifts for 11 of these
galaxies, a redshift success rate of about 45\% (11/24).  These
redshifts are listed in table~\ref{table:specz}.  In all
cases, the redshifts are secured on the basis of the [\ion{O}{2}]
emission line.  We inspected the 2D reduced data and verified that the
emission feature is present in both \textit{independent} spectra
obtained from the nod--and--shuffle observation.     All of these
galaxies  were targeted as candidate members of this galaxy cluster.
Seven of the galaxies have $1.62 < z < 1.63$ with a mean $\langle z
\rangle = 1.622$.   Of the remaining galaxies, two have $z=1.649$ and
1.609 (velocity separations of $<$3100 km s$^{-1}$), and  the other two have
$z=1.536$ and $z=1.496$.  Including all nine galaxies within 3100
km s$^{-1}$, we derive a mean redshift $\langle z \rangle = 1.625$.    

\begin{figure*}
\epsscale{1.04}  
\plotone{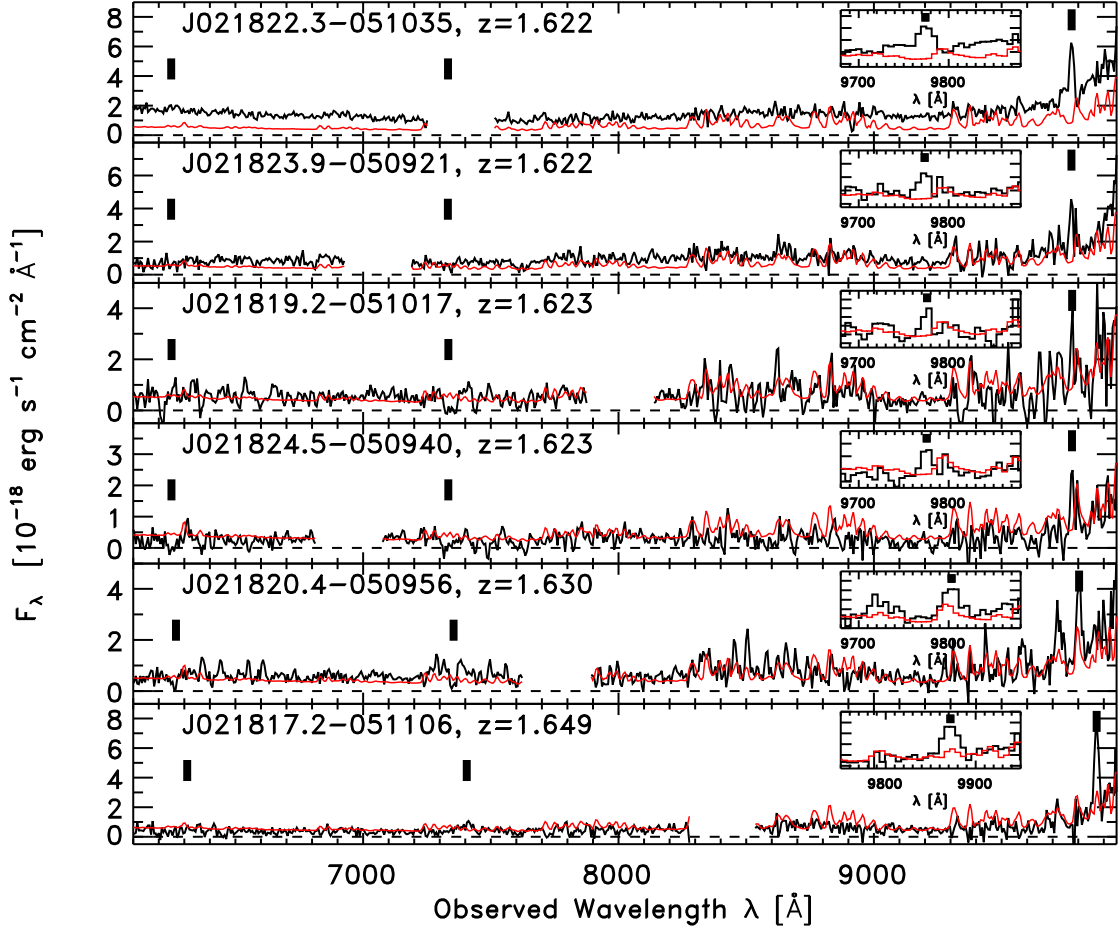}
\epsscale{1.0}
%
\caption{ Extracted IMACS spectra for the six galaxies with $1.62 < z
< 1.65$ within 1.4\arcmin\ of the cluster center (corresponding to
$r < 0.7$~Mpc at $z=1.62$).  The inset panel shows the region around
[\ion{O}{2}] $\lambda$3727 at the measured redshift.   In all panels,
the black line is the measured spectrum.  Gaps in the spectra show
wavelengths that fall on the ``gaps'' between the IMACS CCDs.   The
red line shows the 1$\sigma$ uncertainties propagated through the data
reduction. The thick black lines show the expected locations of \ion{Fe}{2}
$\lambda 2374$, \ion{Mg}{2} $\lambda2800$, and [\ion{O}{2}]
$\lambda3727$ for the measured redshift.\label{fig:spec1d}}
\end{figure*}

Figure~\ref{fig:spec1d} shows the one--dimensional spectra for six
galaxies with $1.62 < z < 1.65$ with the smallest angular separation
($\Delta < 1\farcm4$)  from the astrometric center of \target.  Each
spectrum shows the presence of [\ion{O}{2}] in emission.  Several of
the spectra show evidence for interstellar absorption from \ion{Mg}{2}
and possibly \ion{Fe}{2} (one galaxy at $z=1.649$ shows weak
\ion{Mg}{2} emission, likely from an AGN).   

\begin{figure}
\epsscale{1.17}  
\plotone{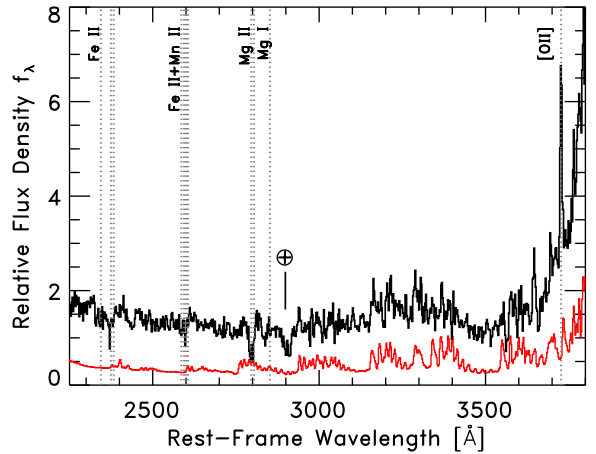}
\epsscale{1.0}
%
\caption{Composite one--dimensional IMACS spectra as a function of
rest--frame wavelength for the seven galaxies within $r < 1.4$~Mpc of
the astrometric center of \target\ with spectroscopic redshifts $1.620
\leq < z \leq 1.630$ (see Table~\ref{table:specz}).     The black line
shows the composite, weight--averaged spectrum, and the red line is
the weighted uncertainty, dominated by emission from the sky.  Strong
[\ion{O}{2}] is observed, as well as several absorption features,
including the doublet \ion{Mg}{2}$\lambda\lambda 2976$,2804, and
absorption from blends of  \ion{Fe}{2} line.  The absorption feature
at rest--frame 2900~\AA\ corresponds to telluric absorption at
$\approx$7600~\AA\ in the observed frame.}  \label{fig:speccomp}
\end{figure}

As the rest of the absorption features are weak, we show in
figure~\ref{fig:speccomp} a weighted--mean, composite spectrum for the
seven galaxies with $1.62 < z < 1.63$, all within $r < 1.4$~Mpc.   To
construct this spectrum, we scaled each spectrum to the mean flux
density in the rest--frame wavelength range 2600--3100~\AA.  We then
weighted each spectrum by the inverse variance using the uncertainty
spectrum.  The stacked spectrum shows broad features from
$\sim$3200--3400~\AA, which are artefacts of the stacking procedure.
Because we have scaled each spectrum to the mean flux, the fainter
galaxies contribute more noise to the stack.  These features correlate
with strong sky emission, and a result of the higher noise in this
region of the spectrum.    The stacked spectrum shows a strong upturn
in flux density at the red end of the spectrum.    This feature is
unlikely to be intrinsic to the galaxies, and arises from the fluxing
uncertainties at the red end of the spectra, due to the decreasing
sensitivity of the CCD and the red cut-off of the blocking filter.   Nevertheless, strong
emission from [\ion{O}{2}] $\lambda$3727 is observed in the composite
spectrum, which is a result of the spectroscopic identification
process.  The stacked spectrum shows several strong absorption
features, in particular \ion{Mg}{2} $\lambda\lambda 2976$,2804,
\ion{Mg}{1} $\lambda$2851 and absorption from various blends of
\ion{Fe}{2} lines.   

The narrow redshift range of the galaxies in table~\ref{table:specz}
suggests they are physically associated.  Using the spectroscopic
redshifts to infer the dynamical conditions of the cluster is dubious
because of the unknown dynamical state (see \S~4.2).     If we
assume the galaxies within $<$1500 km s$^{-1}$ of the mean redshift
and within $0.9$~Mpc of the astrometric center of the
IRAC--selected overdensity sample adequately a virialized structure, then their redshifts correspond to a
line--of--sight velocity dispersion of $\sigma = 860\pm490$ km
s$^{-1}$ using the definition in \citet{carl96}.  The uncertainty is
derived from a bootstrap resampling of the data (and we make no
attempt to remove the additional uncertainty from the redshift
errors).   However, we further qualify this velocity dispersion
because of several biases that likely affect the redshift success rate
for galaxies in this cluster.  Firstly, we prioritized only those
galaxies with $R < 23.8$ mag for spectroscopy, preferentially
excluding fainter galaxies.  Second, even with this magnitude
criterion, most of the galaxies are faint ($23 < R < 24$ mag) and all
the redshift identifications above are based primarily on the
[\ion{O}{2}] $\lambda$3727 emission line.  At $z=1.62-1.63$, this line
falls at $9780-9800$~\AA, adjacent to a strong sky line at 9800~\AA,
which hinders the identification of galaxies at $z\approx 1.63$ with
emission lines (the only galaxy at this redshift identified above
includes strong \ion{Mg}{2} absorption).  Therefore, our spectroscopic
redshifts will be biased against objects at $z\approx 1.63$.   If the
cluster lies at $z=1.625$ then roughly half the galaxies with
redshifts in the upper end of the distribution will be preferentially
missed, biasing the mean redshift and the velocity dispersion
measurement.     Therefore, while we conclude the galaxies are
physically associated as a galaxy (proto--)cluster, we caution against
too much interpretation of the dynamics of this cluster using the
redshifts above. 

\subsection{\textit{XMM-Newton} X-ray Imaging}\label{section:xmm}

\begin{figure}
\epsscale{1.15}  
\plotone{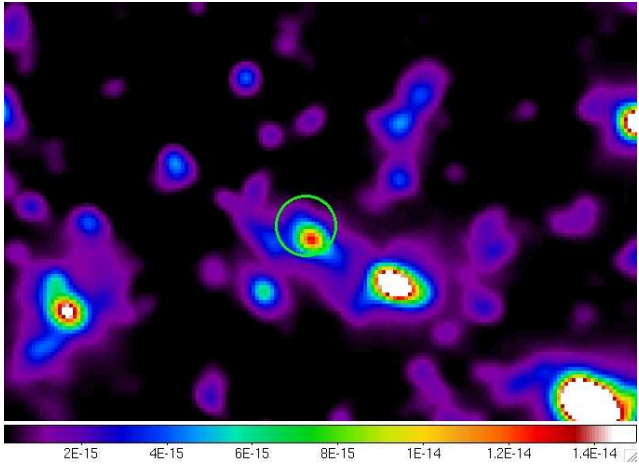}
\epsscale{1.0}
%
\caption{Summed \textit{XMM} image centered on the position of
\target.     \target\ is located on the edges of three adjacent
\textit{XMM} positions.  The coadded \textit{XMM} data yield an
unambiguous source with $4\sigma$ significance in the [0.5--2.5]~keV
band within 10\arcsec\ of the position of the cluster.  The circle
indicates the associated  X-ray source and has a diameter of
40\arcsec.   The units of the color scale is erg s$^{-1}$ cm$^{-2}$
arcmin$^{-2}$ as indicated by the running bar along the bottom of the
image.}
\label{fig:xmm}
\end{figure} 

\target\ is located within the 1$^\circ$ XMM Subaru Deep Field
enclosed in the XMM-LSS survey \citep{pierre04}.    While our cluster
selection method did not involve  X-ray criteria, the
presence of diffuse X-ray emission at the cluster location would
provide independent confirmation of the existence of a deep
gravitational potential well.    In a recent analysis of these data
\citet{fino10} identified two diffuse X-ray sources at distances of
$1\farcm8$ and $2\farcm9$ from \target, with redshift estimates of
$z=1.6-1.8$.   These may be associated with \target\ although their
large offset in angular distance challenges this interpretation.      

Following our discovery of the close association in redshift of the
galaxies in \target, we performed a posteriori analysis of the
\textit{XMM} data at the location of this object.  \target\ is present
on three adjacent \textit{XMM} pointings, near the edge of
each of them (off-axis values $>$12\arcmin).   We coadded the three
\textit{XMM} images centered on the position of \target, and applied
an adaptive smoothing.  The X-ray image is shown
in figure~\ref{fig:xmm}.   The image has an effective exposure time of
$\approx 3\times12$ ks at the position of \target.  There is an
unambiguous source within a radius of 10\arcsec\ of the cluster
position detected with a $4\sigma$ significance.    Within a
40\arcsec--diameter aperture we measured $65$ photons ($\pm 30$\%) in
the [0.5--2~keV] band.   Owing to the  low number of photons and the
large off-axis position of \target\ in the \textit{XMM} images, we are
unable to constrain accurately the spatial extent of the X-ray source
nor to estimate a gas temperature.  However, the detection of the X-ray
emission provides an estimate for the virial mass of \target, which we
discuss in \S~4.2. 
%

\section{THE NATURE OF \target}

In this section we discuss the evidence that \target\ is a galaxy
(proto--)cluster.   Strictly speaking, a
``cluster'' is an object that is fully virialized, while a
``proto--cluster'' is an object that will eventually virialize at
later times (lower redshift).  It remains to be seen if the
dark--matter halo of \target\ is fully virialized or if it is still
assembling.   Nevertheless,  we will blur the distinction between these
definitions and use the term ``cluster'' to mean both.

The photometric redshift distributions and spectroscopic redshifts of
galaxies in \target\  indicate a large overdensity of galaxies at
$z=1.62$ with a high surface density within $r < 0.7$ Mpc.  This is
illustrated in figure~\ref{fig:rgb}.    The galaxies with
spectroscopic redshifts in this region of high surface density show a
preponderance of sources around $z=1.62$.       Furthermore, the
IRAC--selected galaxy members of this cluster have high integrated
redshift probability as defined in equation~1, implying a high
likelihood of being at the cluster redshift.   Galaxies with
$\mathcal{P}_z > 0.5$ are indicated with circles in
figure~\ref{fig:rgb}.  This includes many red galaxies, which are
concentrated near the center of \target\ and have very high
$\mathcal{P}_z$ values.

\begin{figure}
\epsscale{1.15}
\plottwo{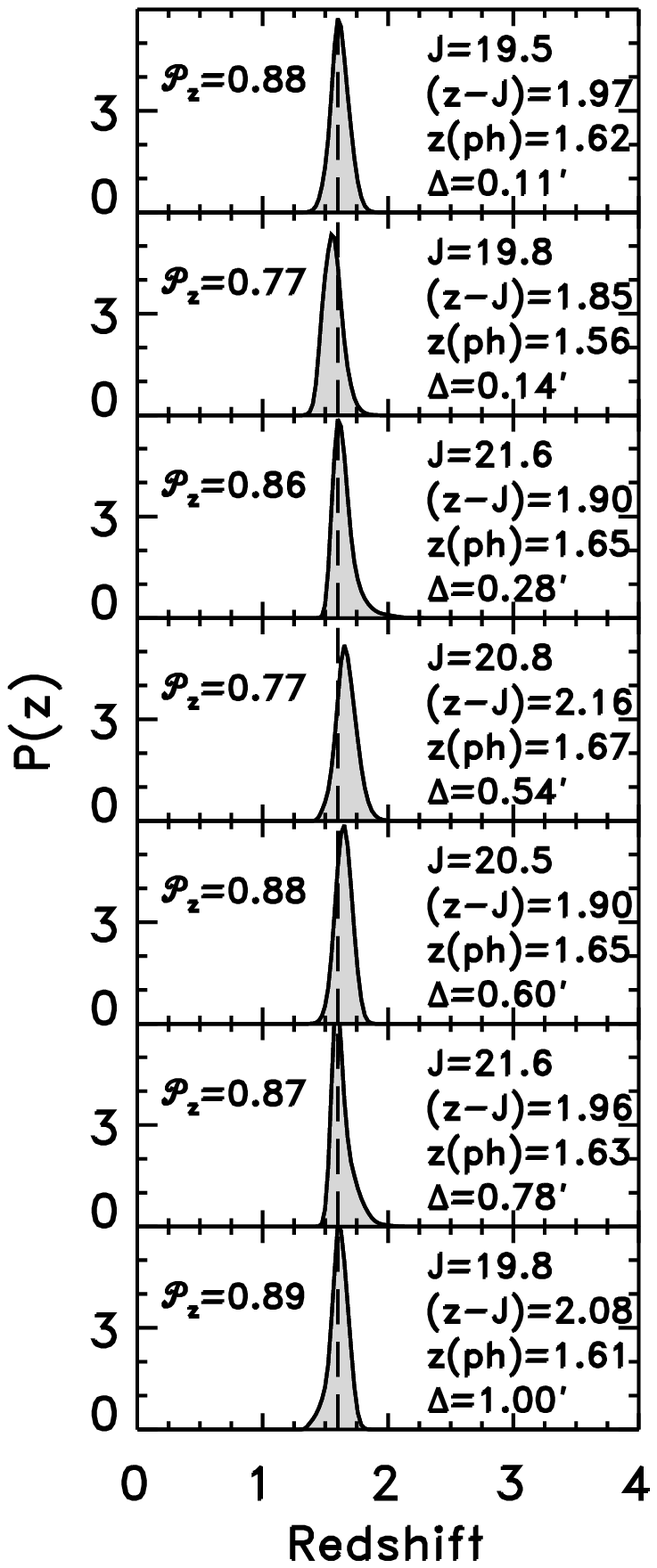}{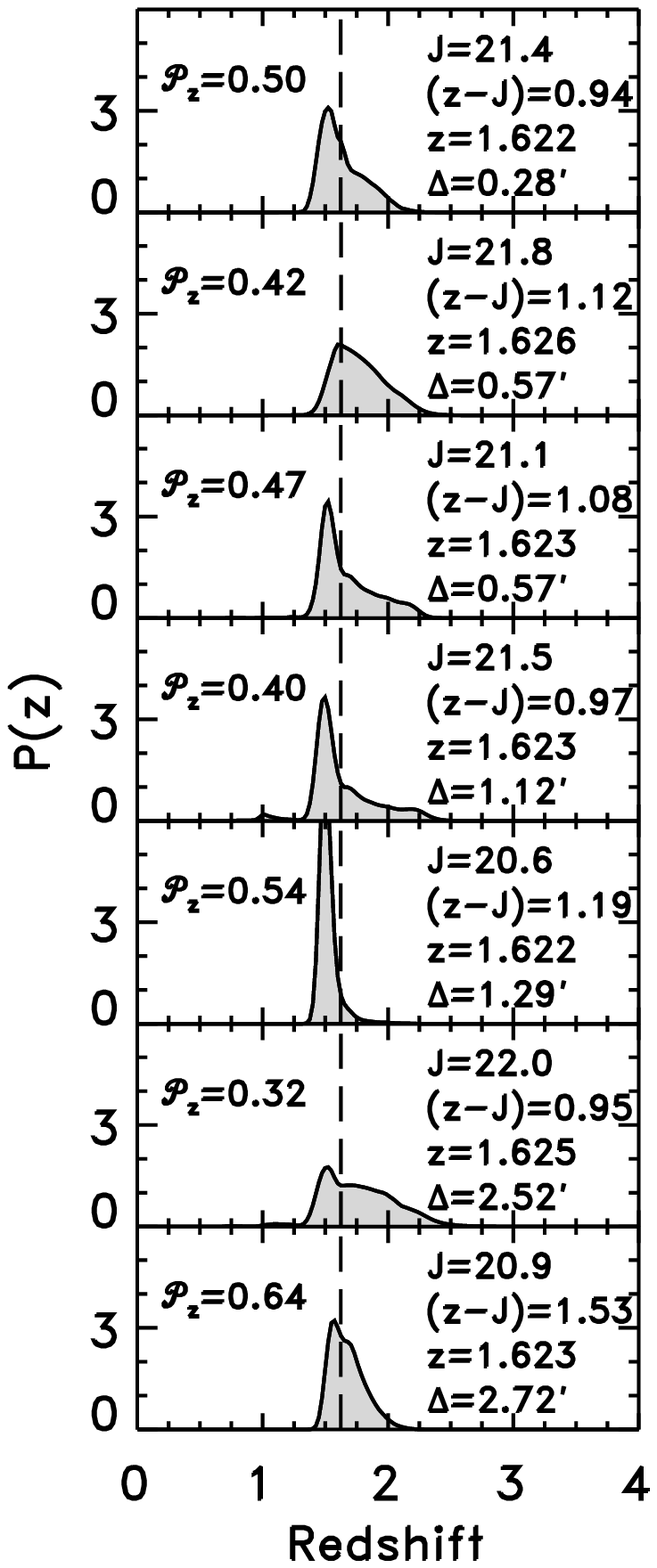}
\epsscale{1.0}
\caption{Photometric--redshift probability--distribution functions,
$P(z)$, for galaxies in \target.   The left figure shows the $P(z)$
for the seven galaxies that have integrated redshift probabilities,
$\mathcal{P}_z > 0.75$ with angular separations of $\Delta <
1$\arcmin\ of the cluster center.   Without exception, these galaxies
all have very red $(z-J)$ colors.  Each panel gives the $(z-J)$ color,
$J$-magnitude, the most likely photometric redshift, the integrated
redshift probability,  and angular separation from the cluster center.
Although these galaxies lack spectroscopic information, their large
integrated redshift probabilities support the assertion that they are
at the cluster redshift, $z=1.62$, indicated by the vertical dashed
lines.    The cluster galaxies with spectroscopic redshifts all show
[\ion{O}{2}] in their spectra, implying ongoing star formation (see
figure~\ref{fig:spec1d}).   They have bluer $(z-J)$ colors and lower
values of $\mathcal{P}_z$, presumably because they have weaker
4000~\AA/Balmer breaks owing to the star formation activity.  The
right figure shows the $P(z)$ for the seven galaxies with
spectroscopic redshifts $1.62 < z < 1.63$.  Each panel shows the same
information as for the left figure, except that they give the
spectroscopic redshift instead of the most likely photometric
redshift. }\label{fig:pz}
%
%
\end{figure}
 
Figure~\ref{fig:pz} shows a montage of photometric redshift
probability distribution functions, $P(z)$, for the seven galaxies
with the
highest integrated redshift probabilities, $\mathcal{P}_z$, with an
angular separation of $\Delta < 1$\arcmin\ of the cluster center.
Without exception these galaxies have red $(z-J)$ colors.    We
targeted several of these galaxies with spectroscopy from IMACS during
our 2009 observing run.  However, their spectra show only faint
optical continua, with no discernible emission features.
Nevertheless, while we are unable to derive spectroscopic redshifts
for these galaxies, the lack of any emission features is consistent
with their photometric redshifts.      The photometric redshifts of
these galaxies are driven by the strength of the apparent
4000~\AA/Balmer break redshifted between the $z$- and $J$-bands at
$z=1.62$.   As a result, the photometric redshift $P(z)$ functions are
sharply peaked at this redshift, and these galaxies have
$\mathcal{P}_z > 0.75$, implying that more than 75\% of their redshift
probability lies between $1.49 < z < 1.76$ with a most likely redshift
at $z\simeq 1.62$, the mean spectroscopic redshift.    

In contrast, the galaxies for which we obtained spectroscopic
redshifts have lower $\mathcal{P}_z$.  These galaxies all show
[\ion{O}{2}] in their spectra, implying relatively recent star
formation.  Therefore they have weaker 4000~\AA/Balmer breaks and
subsequently less-well constrained $P(z)$.   This is illustrated in
figure~\ref{fig:pz}, which shows the $P(z)$ for the seven galaxies
with spectroscopic redshifts at that of the cluster.  All of these
galaxies have bluer $(z-J)$ colors, consistent
with the evidence of more recent star formation.  

\begin{figure*}
\epsscale{1.}  
\plotone{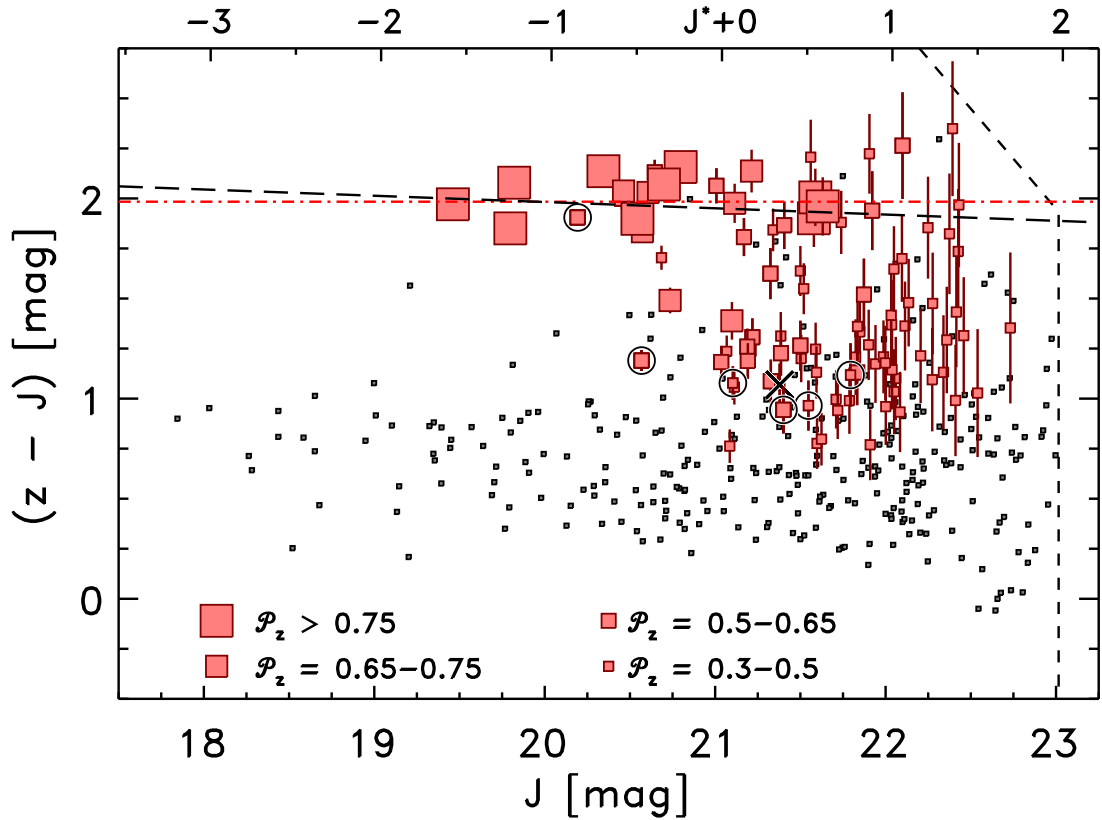}
\epsscale{1.0}
\caption{ Color--magnitude diagram of all galaxies within $\Delta <
2\arcmin$ of the cluster center, corresponding to $1$~Mpc
at $z=1.62$.  The top axis shows magnitudes in units of $J^\ast$, the
characteristic luminosity evolved from the measured values for the
Coma cluster \citep{deprop99} to $z=1.62$. The symbol size denotes
that likelihood that the galaxy lies at the cluster redshift based on
the integrated redshift probability, $P_z$, 
%
%
 as indicated in the legend.    The smallest gray squares all have
$\mathcal{P}_z < 0.3$, and are likely foreground or background field
galaxies, unassociated with the cluster.     Larger, red squares
correspond to galaxies with $\mathcal{P}_z > 0.3$, and likely
associated with the cluster.  The long--dashed line shows a fit to the
red sequence (see \S~4).  The red dot-dashed line shows the expected
color of a stellar population observed at $z=1.62$ formed in a short
burst at $z_f = 2.4$ as described in the text.  Circles denote those
objects spectroscopically confirmed from $1.62 < z < 1.65$.  These are
bluer galaxies owing to the presence of [\ion{O}{2}] emission,
facilitating their spectroscopic identification (see \S~\ref{section:specz}).  The X
symbol shows an object with a spectroscopic redshift outside this
range.   The short dashed lines indicate the $5\sigma$ magnitude limits
derived for point sources in the imaging data.  A strong red--sequence
is apparent, which is dominated by galaxies with high
$\mathcal{P}_z$. 
 \label{fig:cmd}}
\end{figure*}

The red galaxies in this overdensity dominate the color--magnitude
relation for this cluster, similar to the relations observed
in clusters at $z \lsim 1.4$
\citep[\eg,][]{bower92,ellis97,stan98,vandokkum98a,vandokkum98b,blak03,blak06,delucia07b,tran07,lidman08,mei09}. Figure~\ref{fig:cmd}
shows a $z-J$ versus $J$ color--magnitude diagram for \textit{all}
galaxies within 2 arcmin ($r < 1$~Mpc at $z=1.62$) of the
astrometric center of the IRAC--selected overdensity.  The $z$ and $J$
bandpasses span the redshifted 4000~\AA/Balmer break at redshift
$z=1.62$ providing strong contrast between the cluster galaxies and
those in the field.  Field galaxies are shown as small data points in
figure~\ref{fig:cmd}, and these have $\mathcal{P}_z < 0.3$, implying
they are foreground or background galaxies.  Larger symbol sizes
correspond to higher values of $\mathcal{P}_z$, implying those objects
have a greater likelihood of being at the cluster redshift.    The
cluster galaxies on the red--sequence typically have the highest
values of $\mathcal{P}_z$, especially at the bright end where
photometric uncertainties have a smaller effect on the colors.   The
well--defined color--magnitude relation in \target\ is evidence for
the cluster--like nature of this object.  This is the highest redshift
spectroscopically confirmed cluster with such a well--defined red
sequence. 

The red galaxies that dominate \target\ are centrally concentrated,
and nearly all are within $\Delta < 1\arcmin$ of the cluster center.
Very few of the galaxies on the red sequence have angular separations
$>$1\arcmin.  Of the few that have larger angular separations, most
reside within the surface--density contours that extend slightly to the
north--east and north--west from the central region (see
figure~\ref{fig:rgb}).  This property of \target\ is consistent with
the color--density relations observed in other clusters at low and
high redshifts
\citep[\eg,][]{dres80,lidman08, mei09} providing further support for the
cluster--like nature of \target.

In summary, the spectroscopic and photometric redshifts, the
color--magnitude relations, and the surface density--color relations
all support the conclusion that \target\ is a galaxy cluster.  This
interpretation is reinforced by the faint X--ray emission measured at
the location of this object (see also \S~4.2).   We therefore conclude
these galaxies are physically associated with each other as a cluster,
even though we do not have the data to determine if they are fully
virialized.  In the next section we discuss the properties of this
object and the cluster galaxies. 
 
\section{DISCUSSION}

From the evidence presented above, we conclude that \target\
represents a galaxy cluster at $z=1.62$.  Although other
candidates for high--redshift galaxy (proto--)clusters at $z > 1.5$
have been reported
\citep[\eg,][]{miley04,brod07,mccarthy07,zirm08,eise08,andr09,kurk09,chia10},
this is the highest redshift, spectroscopically--confirmed cluster
with a strong, well--defined red sequence.     While the definition of
galaxy ``cluster'' may be subject to semantics (see above), \target\
shows all the characteristics indicative of lower--redshift, rich
galaxy clusters.  

The fact that \target\ has a well--defined red sequence of strongly
clustered red galaxies is not a result of the selection method.
\target\ was selected as an overdensity of galaxies with red IRAC
colors (see \S~2.1).  The IRAC colors at this redshift are blind
largely to variations of  rest--frame optical colors, and should be
sensitive to galaxies with both rest--frame red and blue UV--optical
colors \citep[see][]{papo08}.  With our full
sample, we will compare the properties of \target\ against other
 overdensities of galaxies at this redshift to determine how
the galaxies in this object compare to other co--eval
(proto--)clusters and to field galaxies.  

\subsection{Color Evolution and Formation Epoch}

As discussed above, \target\ is dominated by a strong red sequence of
bright galaxies.  These are intrinsically very luminous for their
redshift, $z=1.62$.  The top axis of figure~\ref{fig:cmd}
compares the measured $J$--band magnitudes to the ``characteristic''
magnitude, $J^\ast$, for the luminosity function of galaxies in the
Coma cluster, evolved passively backwards in time to $z=1.62$ using
the model of \citet{deprop99}.  The brightest galaxies in \target\
correspond to $J^\ast - 1$ to $J^\ast - 1.5$~mag.  The descendants of
these galaxies at a minimum will be super--$L^\ast$ cluster galaxies by
$z\sim 0$, even without subsequent merging or star formation. 
%
%

The red galaxy colors imply that they contain older stellar
populations.  Figure~\ref{fig:cmd} illustrates the expected color of a
composite stellar population formed at $z_f = 2.4$ with a
star--formation rate that evolves by an  e--folding timescale $\tau =
0.1$~Gyr to $z=1.62$ using the 2007 version of the \citet{bruz03}
models.  This model produces a $(z-J)$ color similar to what we
measure for the brighter red galaxies in this cluster (using the
Bruzual \& Charlot 2003 models produces $z-J$ colors within $0.02$~mag
for these ages and formation redshift).   There is also no indication
that \target\ contains a population of blue, luminous galaxies.  The
brightest galaxies within a projected separation of $r < 1$~Mpc of
\target\ with $(z - J) < 1.5$~mag and $\mathcal{P}_z > 0.3$ have
fainter magnitudes, typically sub--$J^\ast$.  Because the
mass--to--light ratios of the stellar populations increase as they
redden and fade over time, the dominant population of red--sequence
galaxies could not form directly from the less--luminous blue
galaxies.    Interestingly, the number of red sequence galaxies
appears to decline toward the faint end of the red sequence, seemingly
at magnitudes well above the detection limit (see
figure~\ref{fig:cmd}).   This observation is similar to the findings
of \citet{tanaka07}, who observed a deficit of red galaxies around a
$z=1.24$ cluster.     These results imply a strong evolution in the
faint end of the red--sequence luminosity function, extending the
relation measured by \citet{rudn09} for moderate redshift clusters to
higher redshift.   It also appears that there is a lack of faint blue
galaxies with $\mathcal{P}_z>0.3$ but no such lack is seen in the
$\mathcal{P}_z < 0.3$ galaxies, indicating that the photometry is
indeed complete to the level indicated in the figure.  Further studies
of clusters at these redshifts and detailed modeling are  needed to
confirm this.
%

\begin{figure}
\epsscale{1.2} \plotone{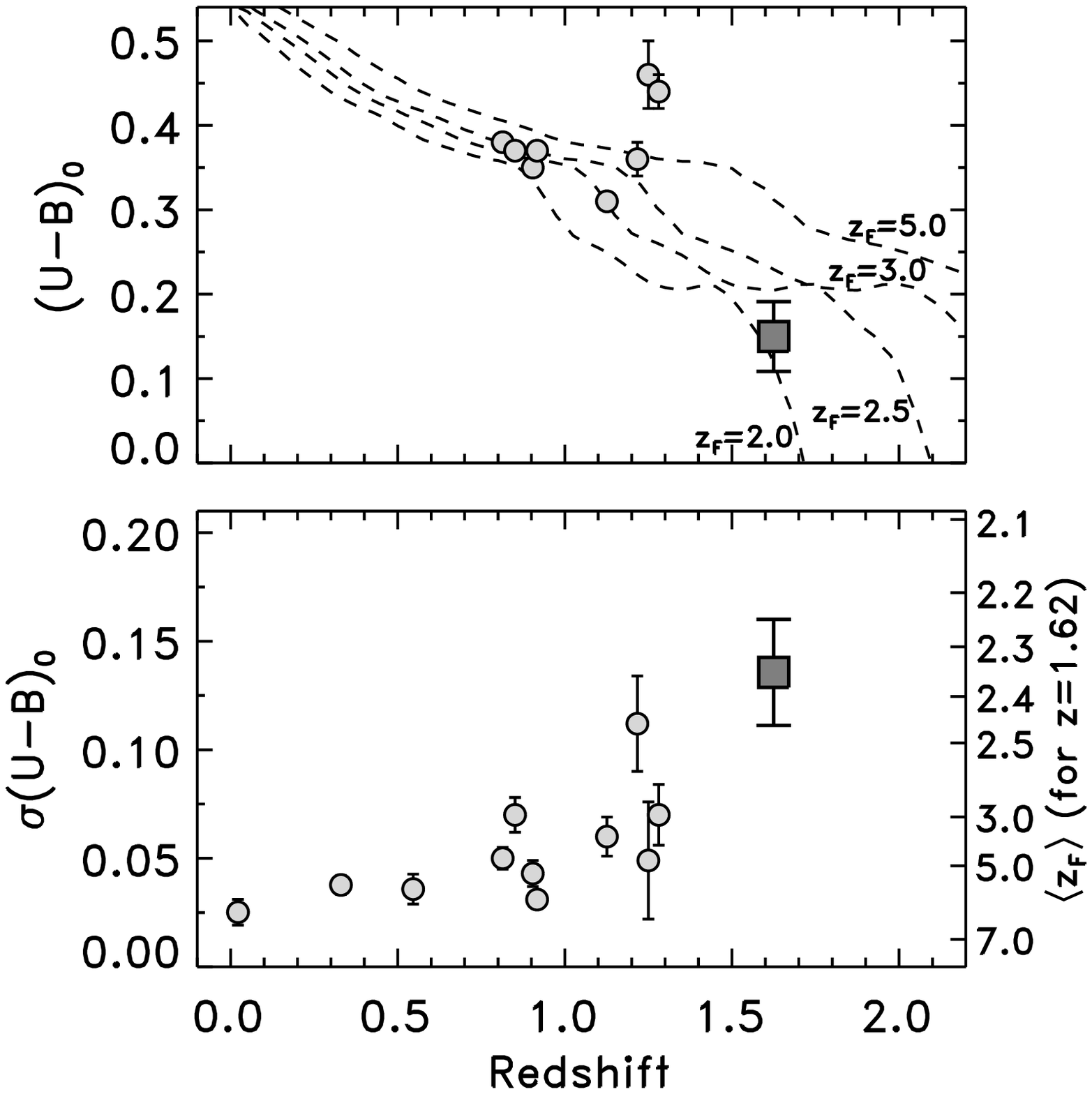}
\epsscale{1.0} 
\caption{  The top panel shows the evolution of the rest--frame $U-B$
color measured at rest--frame $M_{B}=-21.4$~mag for red--sequence
galaxies in clusters as a function of redshift.  Smaller circles show
the results from lower--redshift clusters
\citep{bower92,ellis97,vandokkum98a,vandokkum98b,vandokkum00,mei09}.
The large box point shows the value derived for the cluster \target\
at $1.625$ derived here.   The curves show the expected evolution of a
stellar population with solar metallicity formed with an e--folding
timescale $\tau=0.1$~Gyr at the formation redshift indicated.     The
bottom panel shows the evolution of the scatter in the rest--frame
$U-B$ color for red--sequence galaxies as a function of redshift.
Symbols are the same as the top panel.   The scatter in the colors of
red--sequence galaxies increases with redshift.    The right axis of
the bottom  panel indicates the expected scatter in the rest--frame
$(U-B)$ color for a red sequence observed at $z=1.62$ for a given mean
luminosity--weighted formation redshift (see text).  The scatter in
the $(U-B)$ color for the $z=1.62$ cluster corresponds to a
luminosity--weighted formation redshift of $\bar{z_f} = 2.25-2.45$,
consistent with the evolution in the $(U-B)_0$
color.  \label{fig:umb}}
\end{figure}

The zeropoint, scatter, and slope of the red sequence itself provide
constraints on the formation timescales of the galaxies' stellar
populations  \citep[\eg,][]{bower92,aragon93}.   To compare the
galaxies in \target\ at $z=1.62$ to those in lower redshift clusters,
we converted the observed $(z - J)$ colors to rest--frame $(U-B)$
colors following \citet{mei09}.  We find no evidence for evolution in
the slope of the red sequence between \target\ and lower redshift
galaxy clusters, consistent with the results of other studies
\citep[see, \eg,][]{mei09}.   The top panel of figure~\ref{fig:umb}
shows the rest--frame $(U-B)$ color of \target\ and lower redshift
clusters taken from the literature at a fixed rest--frame absolute
magnitude, $M_{B} = -21.4$~mag.    However, there is strong evolution
in the rest--frame $(U-B)$ colors, which become redder with decreasing
redshift, consistent  with passive evolution.  The models in
figure~\ref{fig:umb} illustrate the expected evolution for stellar
populations with different formation redshifts ($z_F = 2.0$, 2.5, 3.0,
and 5.0), with a star--formation e--folding timescale, $\tau =
0.1$~Gyr, and solar metallicity.     The $(U-B)$ color at $M_B =
-21.4$~mag for \target\ is consistent with stellar populations formed
between $2.0 \lsim z \lsim 2.5$.   

We model the scatter in the colors of the red--sequence galaxies in
\target\ by following \citet[and references therein]{hilton09}.  We
construct a series of composite stellar populations  with
star--formation e--folding timescale of $\tau = 0.1$~Gyr and solar
metallicity.  We tested other values for the metallicity (ranging from
0.2 $Z_\odot$ and 2.5 $Z_\odot$), but found that they did  not
reproduce simultaneously the zeropoint and scatter in the rest--frame
$(U-B)$ colors for a consistent formation epoch.   In our model each
galaxy forms stars in a single burst of star formation that begins at
an initial formation redshift, $z_F$, and lasts for a duration $\Delta
t$.   We allow the  duration $\Delta t$ to vary from $\Delta t=0$ to a
maximum equal to the lookback time from $z_F$ to $z=1.62$, the
redshift at which the cluster is observed.  We allow for a range of
formation redshift (with a maximum formation redshift taken to be when
the lookback time was 4 Gyr), where at each, $z_F$, we construct a
simulated sample of $10^5$ galaxies with ages assigned randomly from a
uniform distribution with $0 < t < \Delta t$.  From these simulated
galaxies we compute the scatter in the simulated color distributions
and the luminosity--weighted age.   Then, for a given measurement of
the scatter in the colors,  we infer the corresponding
luminosity--weighted age and thus obtain an estimate for the formation
redshift of the red--sequence galaxies in the cluster.

We calculated the scatter for the red sequence galaxies in \target,
including those galaxies with integrated redshift probability
$\mathrm{P}_z > 0.3$ within a projected radius of 1 Mpc (2 arcmin) of
the cluster center.    We used an interative rejection algorithm to
include only red galaxies within $2.5\sigma$ of the derived red
sequence.  We derive the scatter about the median absolute deviation
\citep{beers90}, which yields $\sigma(U-B)  = 0.136\pm
0.024$.   We compare this to the red
sequences of other galaxy clusters  and our models in the bottom
panel of figure~\ref{fig:umb}.       There is a marked increase in the
scatter in the $(U-B)$ color with redshift, which is expected because
the cluster
galaxies are observed closer in time to when they formed their stellar
populations \citep[\eg,][]{mei09,hilton09}.  The right axis of the
panel indicates the modeled relationship between $\sigma(U-B)$ and the
formation redshift, $z_f$ for the model stellar populations observed
at $z=1.62$.  The scatter in the rest--frame $(U-B)$ color for
\target\ corresponds to range of formation redshift, $2.25 \leq z \leq
2.45$, using the 2007 version of the \citet{bruz03} stellar population
synthesis models (using the 2003 models has a negligible change on the
formation redshifts for the stellar population ages involved here).
We note that this is an upper limit as we have made no attempt to
remove the color measurement uncertainties from the intrinsic scatter.
If we have overestimated the scatter, the formation redshift will be
\textit{higher}.  Given the agreement between the formation epochs
derived from the red-sequence zeropoint and scatter, we do not think
this is a serious effect.  

Therefore, both the zeropoint and scatter of the rest--frame $(U-B)$
colors for red--sequence galaxies in \target\ imply a formation epoch
of $z_f \simeq 2.25-2.45$ (a lookback time of $1.0-1.3$~Gyr from
$z=1.62$).  This corresponds to the last major star--formation episode
in these red cluster galaxies.  This is a similar formation epoch as
found in many studies of other $z>1$ cluster galaxies
\citep[\eg,][]{mei09,hilton09,rosati09}, although at least one cluster
(XMM J2235.3-2557 at $z=1.39$) has galaxies with formation epochs
ranging from $z_f\sim 2$ to $z_f\sim 6$ \citep{rosati09}.  The
formation epoch of \target\ is also consistent with the colors and
spectral indices of galaxies in lower--redshift massive clusters
\citep[\eg,][]{stan98,vandokkum07a}.  Furthermore, studies show that a
high fraction of massive galaxies at $z > 2$ have high levels of star
formation \citep[\eg,][]{kriek06,papo06a}, and these are likely
consistent with the evolution of the cluster galaxies.  These
properties all support the conclusion that the galaxies of \target\
will become typical galaxies found in rich clusters at later times. 

\subsection{Dynamical Mass Estimate}

As discussed above, it is unclear if the cluster galaxies of \target\
sample a virialized, relaxed dark--matter halo, or if this object is
in the process of assembling through mergers.   Under the assumption
that the galaxies are virialized, we use the velocity dispersion to
provide a crude estimate  for the dynamical mass of \target\ as a
reference, although we qualify this with the caveats in \S~\ref{section:specz}.  For a velocity dispersion, $\sigma=860\pm 490$~km s$^{-1}$
(see \S~\ref{section:specz}), the  corresponding dynamical mass estimate is
$M_\mathrm{dyn}  \sim 3 r_\mathrm{vir} \sigma^2 \sim 4 \times
10^{14}$~\msol\ for a virial radius of 0.9~Mpc.  We
derive the same dynamical mass estimate using $M_{200}$
\citep{carl97}.\footnote{Here $M_{200}$ is the mass of the dark matter
halo within a spherical volume defined by radius $r_{200}$ where the
average density is $200$ times the critical density at the observed
redshift.}   Even so, we caution the reader that the dynamical mass
may be highly uncertain given the statistical errors and systematic biases. 

This mass is a factor of $\sim$4 larger than the limiting halo mass of
the IRAC--selected clusters \citep{papo08}, and if accurate then it
implies that \target\ resides in one of the most overdense
environments in the SWIRE survey.     In this case, based on the
expected growth of dark--matter haloes \citep{spri05a} the mass of
\target\ should increase to at least $10^{15}$~\msol\ by $z=0.2$,
becoming a rich cluster of galaxies comparable to the Coma cluster.  

The high dynamical mass estimate of \target\ corresponds to a large
gravitational potential well.    If this is the case, then the hot diffuse
gas of the ICM should emit significant X-ray
luminosity.    This is supported by the X--ray detection we infer at
the location of \target\ in \S~\ref{section:xmm}.  Assuming that all
the detected X--ray emission originates from the ICM we infer a mean
temperature of $\approx$3~keV and virial mass of $1.1\times
10^{14}$~\msol\ based on the local luminosity--temperature and
mass--temperature relations \citep{arnaud99,arnaud05}.  This is
roughly a factor of 4 lower than the dynamical mass derived above, but
within the large error budget.    While the X--ray detection of
\target\ supports the high derived virial mass,  obtaining much deeper
X--ray data is required for a self-contained analysis of the ICM
properties and mass determination.

\section{SUMMARY}

We report the discovery of a spectroscopically confirmed galaxy
cluster at $z$=1.62 located in the SWIRE XMM-LSS field. This cluster
candidate was selected solely as an overdensity of sources with red
\spitzer/IRAC colors, satisfying $(\mone - \mtwo)_\mathrm{AB} > -0.1$
mag, with no other selection criteria imposed. Photometric redshifts
derived from SXDS ($BViz$-bands), UKIDSS-UDS ($JK$-bands), and SpUDS
(3.6-8.0 micron) for the galaxies in and around this cluster show that
this structure corresponds to a galaxy surface density of sources at
$z=1.6$ that is $>20\sigma$ times the mean surface density at this
redshift.    Furthermore,  our recent analysis of existing
\textit{XMM} data on this cluster provides a weak but unambiguous
detection compatible with the expected thermal emission from such a
cluster.  This confirms that our selection of overdensities of sources
with red IRAC colors identifies galaxy clusters at $z \ge 1.3$.  

We obtained spectroscopic observations of galaxies in the cluster
region using IMACS on the Magellan telescope.  We measured redshifts
for five galaxies in the range $z$=1.62--1.63, all within 1.4 arcmin
($< 0.7$~Mpc) of the cluster center.  In addition, we measured
spectroscopic redshifts for two sources with $z$=1.62-1.63 within
$1.4-2.8$ arcmin ($0.7-1.4$~Mpc). The cluster appears to be dominated
by red galaxies, with $(z - J) > 1.7$ mag. The photometric redshift
distributions for the brightest red galaxies are centrally peaked at
$z=1.62$, coincident with the spectroscopically confirmed galaxies. 

The $z-J$ versus $J$ color--magnitude diagram of the galaxies in this
cluster shows a strong red-sequence, which includes the dominant
population of red galaxies. The rest--frame $(U-B)$ color and scatter
of galaxies on the red-sequence are consistent with a mean
luminosity--weighted age of $1.0-1.3$ Gyr, yielding a
formation redshift $\overline{z_f} = 2.35 \pm 0.10$, and
corresponding to the last major episode of star formation in these
galaxies.

We provide a crude estimate of the dynamical mass of \target, although
this result is highly uncertain due to systematics in the data and
owing to the unknown  dynamical state of \target.   Under the
assumption that the dark matter halo of \target\ is virialized and
relaxed, we estimate a dynamical mass based on the measured velocity
dispersion, $M_\mathrm{dyn} \sim 4 \times 10^{14}$~\msol\ within a
virial radius of $r_\mathrm{vir} \sim 0.9$~Mpc.    Our reanalysis of
the \textit{XMM} X-ray emission of this cluster favors a lower virial
mass, $1.1\times 10^{14}$~\msol,  but consistent within the
uncertaintites.    If these masses are accurate, then we expect
\target\ to evolve to a rich cluster with $M \sim 10^{15}$~\msol\ at
$z=0.2$  similar to the Coma cluster.  Further spectroscopic and
multiwavelength observations of galaxies in \target\ are needed to
constrain better  the mass measurement. 

\acknowledgments

We acknowledge our colleagues for stimulating discussions, and helpful
comments.   In particular, we wish to thank Chris Simpson and Masayuki
Akiyama for their spectroscopic redshifts in advance of publication.
We thank Renbin Yan for assistance with the spectroscopic reductions.
We also acknowledge Stefano Andreon, Chris Simpson,
S. Adam Stanford, and Jon Willis for very helpful feedback and
suggestions.  Lastly, we thank the anonymous referee whose helpful
comments improved both the quality and clarity of this paper.  This
work is based on observations made with the \textit{Spitzer Space
Telescope}, which is operated by the Jet Propulsion Laboratory,
California Institute of Technology.  This work is based in part on
data obtained as part of the UKIRT Infrared Deep Sky Survey.  A
portion of the Magellan telescope time was granted by NOAO, through
the Telescope System Instrumentation Program (TSIP). TSIP is funded by
NSF.  Support for M.B. was provided by the W. M. Keck Foundation.  We
acknowledge generous support from the Texas A\&M University and the
George P.\ and Cynthia Woods Institute for Fundamental Physics and
Astronomy.



\end{document}